\documentclass[12pt]{article}

\usepackage{amsmath}
\usepackage{caption}
\usepackage{subfig}
\usepackage{graphicx} 
\usepackage{amsmath}
\usepackage{graphics} 
\usepackage{anysize}
\usepackage{multirow}
\usepackage{wrapfig}
\usepackage{float}
\usepackage[usenames]{color}
\usepackage[hidelinks]{hyperref}

\usepackage{cleveref}
\usepackage{scicite}
\usepackage{times}

\newenvironment{sciabstract}{%
\begin{quote} \bf}
{\end{quote}}

\newcounter{lastnote}

\title{Molecular Identification from AFM images using the IUPAC Nomenclature and Attribute Multimodal Recurrent Neural Networks} 

\author
{Jaime Carracedo-Cosme,$^{1,2}$ Carlos Romero-Mu\~{n}iz,$^{3}$ \\ Pablo Pou,$^{2,4}$ Rub\'en P\'erez,$^{2,4,\ast}$\\
\\
\normalsize{$^{1}$Quasar Science Resources S.L.,}
\normalsize{Camino de las Ceudas 2, E-28232 Las Rozas de Madrid, Spain}\\
\normalsize{$^{2}$Departamento de F\'isica Te\'orica de la Materia Condensada,}\\ \normalsize{Universidad Aut\'onoma de Madrid, E-28049 Spain}\\
\normalsize{$^{3}$Departamento de F\'{i}sica Aplicada I, Universidad de Sevilla, E-41012, Seville, Spain}\\
\normalsize{$^{4}$Condensed Matter Physics Center (IFIMAC),}\\
\normalsize{Universidad Aut\'onoma de Madrid, E-28049 Madrid, Spain}\\
\\
\normalsize{$^\ast$ ruben.perez@uam.es}
}

\date{May 1, 2022}

\usepackage[acronym]{glossaries}
\newacronym{AFM}{AFM}{Atomic Force Microscopy}
\newacronym{CNN}{CNN}{Convolutional Neural Network}
\newacronym{RNN}{RNN}{Recurrent Neural Network or Elman network~\cite{elman1990finding}}
\newacronym{NLP}{NLP}{Natural Languaje Processing}
\newacronym{M-RNN}{M-RNN}{Multimodal Recurrent Neural Network}
\newacronym{LSTM}{LSTM}{Long Short-Term Memory}
\newacronym{GRU}{GRU}{Gated Recurrent Unit}
\newacronym{NN}{NN}{Neural Network}
\newacronym{AMRNN}{AM-RNN}{Atribute Multimodal Recurrent Neural Network}

\newacronym{BLEU}{BLEU}{Bilingual Evaluation Understudy}
\newacronym{relu}{ReLU}{Rectified Linear Unit}
\newacronym{lr}{lr}{Learning Rate}
\newacronym{AFMD}{QUAM-AFM}{Quasar Science Resources S.L. - Universidad Aut\'onoma de Madrid - Atomic Force Microscopy}
\newacronym{Adam}{Adam}{Adaptive Moment estimator}
\newacronym{FC}{FC}{Fully connected}
\newacronym{IDG}{IDG}{Image Data Generator}
\newacronym{Pro-PPM}{Pro-PPM}{Pro Probe Particle Model}
\newacronym{PPM}{PPM}{Probe Particle Model}
\newacronym{RMSProp}{RMSProp}{Root Mean Square Resilient Propagation}
\newacronym{DL}{DL}{Deep Learning}
\newacronym{IUPAC}{IUPAC}{International Union of Pure and Applied Chemistry}
\newacronym{softmax}{Softmax}{Soft Approximation of Max}
\newacronym{DFT}{DFT}{Density Functional Theory}
\newacronym{VASP}{VASP}{Vienna Ab initio Simulation Package}
\newacronym{AI}{AI}{Artificial Intelligence}
\newacronym{M-RNN-AT}{M-RNN$_{A}$}{Multimodal Recurrent Neural Network for attribute prediction}
\newacronym{HRAFM}{HR-AFM}{high resolution AFM}
\newacronym{RES}{RES}{Red Espa\~n{}ola de Supercomputaci\'{o}n}

\begin{document} 


\baselineskip24pt

\maketitle 

\begin{sciabstract}

Despite being the main tool to visualize molecules at the atomic scale, \gls{AFM} with CO-functionalized metal tips is unable to chemically identify the observed molecules. Here we present a strategy to address this challenging task using deep learning techniques. Instead of identifying a finite number of molecules following a traditional classification approach, we define the molecular identification as an image captioning problem. We design an architecture, composed of two multimodal recurrent neural networks, capable of identifying the structure and composition of an unknown molecule using a 3D-AFM image stack as input. The neural network is trained to provide the name of each molecule according to the IUPAC nomenclature rules. To train and test this algorithm we use the novel QUAM-AFM dataset, which contains almost 700,000 molecules and 165 million AFM images. The accuracy of the predictions is remarkable, achieving a high score quantified by the cumulative BLEU 4-gram, 
a common metric in language recognition studies.

\end{sciabstract}

\newpage

\section{Introduction}

Scanning Probe Microscopes have played a key role in the development of nanoscience as the fundamental tools for the local characterization and manipulation of matter with high spatial resolution. 
In particular, \gls{AFM} operated in its frequency modulation mode allows the characterization and manipulation of all kind of materials at the atomic scale~\cite{giessibl1995atomic,PerezSurfSciRep2002,GiessiblRevModPhys2003}. This is achieved measuring the change in the frequency of an oscillating tip due to its interaction with the sample. When the tip apex is functionalised with inert closed-shell atoms or molecules, particularly with a CO molecule, the resolution is dramatically enhanced, providing access to the inner structure of molecules~\cite{gross2009chemical}. 
This outstanding contrast arises from the Pauli repulsion between the CO probe and the sample molecule~\cite{gross2009chemical,Moll2010} modified by the electrostatic interaction between the potential created by the sample and the charge distribution associated with the oxygen lone pair at the probe~\cite{ellner2019molecular,VanDerLitPRL2016,HapalaNatComm2016}. In addition, the flexibility of the molecular probe enhances the saddle lines of the total potential energy surface sensed by the CO~\cite{HapalaPRB2014}. These \gls{HRAFM} capabilities have made possible to visualize frontier orbitals~\cite{GrossAC2018}, to determine bond order potentials~\cite{GrossScience2012} and charge distributions~\cite{GrossScience2009b,MohnNatNano2012}, and have opened the door to track and control on-surface chemical reactions~\cite{deOteyzaScience2013,clair2019controlling}. 

In spite of these impressive achievements~\cite{Giessibl2019review, zhong2020review} one of the most important goals remains elusive: the molecular recognition. That is, the ability of naming a certain molecule exclusively by means of HR-AFM observations.
Molecules have been identified combining AFM with other experimental techniques like scanning tunneling microscopy (STM) or Kelvin probe force microscopy (KPFM), and with the support of theoretical simulations~\cite{GrossAC2018,zhong2020review,HanssenACIE2012,EbelingNatComm2018,Liljeroth2018,SchulerJACS2015}. 
Chemical identification by AFM of individual atoms at semiconductor surface alloys was achieved using reactive semiconductor apexes~\cite{SugimotoNature2007}. In that case, the maximum attractive force between the tip apex and the probed atom on the sample carries information of the chemical species involved in the covalent interaction. However, the scenario is rather different when using tips functionalised with the inert CO molecules where the main AFM contrast source is the Pauli repulsion and the images are strongly affected by the probe relaxation. 
So far, the few attempts to discriminate atoms in molecules by HR-AFM have been based either on differences found in the tip-sample interaction decay at the molecular sites~\cite{ellner2019molecular, HeijdenACSNano2016} or on characteristic image features associated with the chemical properties of certain molecular components ~\cite{ellner2019molecular,zhong2020review,GuoLangmuir2010,SchulerJACS2015,Tschakert2020natcomm,JelinekJPCM2017,GrossAC2018,EXP,zahl2021TMA}. For instance, sharper vertices are displayed for substitutional N atoms on hydrocarbon aromatic rings\cite{GuoLangmuir2010,ellner2019molecular,HeijdenACSNano2016} due to their lone pair. Furthermore, the decay of the CO-sample interaction over those substitutional N atoms is faster than over their neighboring C atoms~\cite{ellner2019molecular,HeijdenACSNano2016}. Halogen atoms can also be distinguished in AFM images thanks to their oval shape (associated to their $\sigma$-hole~\cite{Tschakert2020natcomm}) and to the significantly stronger repulsion  compared to atoms like nitrogen or carbon~\cite{Tschakert2020natcomm}. 
However, even these atomic features depend significantly on the molecular structure~\cite{ellner2019molecular,GrossScience2012} and cannot be only associated to a certain species but  to its moiety in the molecule. The huge variety of possible chemical environments renders the molecular identification by a mere visual inspection by human eyes an impossible task. 



\gls{AI} techniques are precisely optimized to deal with this kind of subtle correlations and massive data. Deep learning (DL), with its outstanding ability to search for patterns, is  nowadays routinely used to classify, interpret, describe and analyze images~\cite{krizhevsky2012imagenet,simonyan2014very, Kaiming2016Residual, szegedy2016inception, chollet2017xception, sandler2018mobilenetv2},  providing machines with capabilities hitherto unique to human beings or even surpassing them in some tasks~\cite{he2015delving}. 
There are two main challenges to apply deep learning to achieve a complete molecular identification (structure and composition) through \gls{AFM} imaging. The first one is the limited amount of experimental data available to train the models. 
In a previous work~\cite{Carracedo2021MDPI},  we have explored the performance of an specifically designed \gls{CNN}, trained with a data set that includes 314,460 theoretical images --calculated with the latest HR-AFM modeling approaches~\cite{ellner2019molecular,liebig2020quantifying}--  and only 540 images generated with a variational autoencoder from very few experimental images. This \gls{CNN}, applied to a set of 60 molecular structures that include 10 different atomic species (C, H, N, P, O, S, F, Cl, Br, I), obtained almost perfect (99\%) accuracy in the classification using simulated  \gls{AFM} images and very good accuracy (86\%) for experimental \gls{AFM} images.  
Encouraged by the success of this proof--of--concept, we have recently extended the available data sets of theoretical \gls{AFM} images with the generation of 
QUAM--AFM~\cite{QUAM-AFM_repository}, that aims to provide a solid basis for making results from DL applications to the AFM field reliable and reproducible~\cite{artrith2021best}.  QUAM--AFM includes calculations for a collection of 686,000 molecules using 240 different combinations of AFM operation parameters (tip--molecule distance, cantilever oscillation amplitude and tilting stiffness of the CO-metal bond), resulting in a total of 165 million images~\cite{QUAM-AFM_repository}.

The second challenge arises from the non--planar structure of the molecules, that mixes up in the molecular charge density --ultimately responsible for the AFM contrast-- the effects of the geometry and the chemical composition, making it very difficult to disentangle them.
Alldritt \emph{et al.}~\cite{alldritt2020automated} developed a \gls{CNN} focused on the task of determining the molecular geometry. Results were excellent for the structure of quasi-planar molecules, even using the algorithm directly with experimental images. For 3D structures, they were able to recover information for the positions of the atoms closer to the tip. However, the discrimination of functional groups produced  non conclusive results. 
At variance with this study, as we already mentioned above, a \gls{CNN}~\cite{Carracedo2021MDPI} was able to solve the classification problem for 60 essentially flat molecules with almost perfect accuracy, being able to identify, for example, the presence of a particular halogen (F, Cl, Br or I) in molecular structures that, apart from this atom, were identical.
Although encouraging, the clear success of this proof of concept does not provide a solution to the general problem of molecular identification.  The classification approach can only identify molecules included in the training data set. Given the rich complexity provided by organic chemistry, even an extremely large data set, that already poses fantastic computational challenges  (as the output vector has the dimension of the number of molecules in the dataset), would fail to classify many of the already known or possibly synthesized molecules of interest. 

In this work, we transform the problem of molecular identification into an image captioning challenge: the description of the content of an image using language. Automatic image captioning has been a field of intensive research for deep learning techniques over the last years~\cite{mao2014deep, you2016image, cornia2020meshed, zhou2020unified}. It has been recently and successfully used \cite{rajan2021decimerv1,Img2MOL} for optical chemical structure recognition~\cite{rajan2020review}, 
the translation of graphical molecular depictions into machine-readable formats. These works are able to predict the SMILES textual representation \cite{SMILES} of a molecule from an image with its chemical structure depiction by using standard encoder-decoder \cite{Img2MOL} or transformer \cite{rajan2021decimerv1} models. In our case, we consider a stack of 10 constant-height HR-AFM images, each corresponding to different tip--sample distances, as the ``image'' and the \gls{IUPAC} name of the molecule as the description or caption.
Most of the current methods for automatic image captioning have two key components: (i) a \gls{CNN} --a \gls{NN} with convolutional kernels as processing units-- that represents the high-level features of the input images in a reduced dimensional space; and (ii) a \gls{RNN} --a \gls{NN} whose units are complex structures that have an inner state that stores the temporal context of a time series-- that deals with language processing and predicts a single word at each time step~\cite{brown1993mathematics,chung2014empirical, sak2014long}. 
The \gls{IUPAC} name determines unambiguously the molecular composition and structure. This is done by defining a hierarchical keyword list to name functional groups that are written following a systematic syntax that defines the structural position of each moiety or group in the molecule~\cite{IUPAC_book}.
Therefore, we tackle the molecular recognition challenge with a deep learning architecture that decomposes into two models. The first one predicts the main chemical groups that compose the molecule whereas the second model performs the \gls{IUPAC} formulation. QUAM--AFM~\cite{QUAM-AFM_repository}, is used to train and test the networks.  Our approach predicts the exact name in almost half of the cases and achieves a high accuracy according to the \gls{BLEU} 
algorithm~\cite{papineni2002bleu}, the most commonly applied metric to score the accuracy of language-involved models. 

\section{A Deep Learning Approach for Molecular Identification}

\subsection{QUAM-AFM: Structures and AFM Simulations}

One of the main challenges to automate the molecular identification through \gls{AFM} imaging arise from the limited availability of data to fit the parameters of deep learning models. We use \gls{AFMD}~\cite{QUAM-AFM_repository}, a dataset of 165 million \gls{AFM} images theoretically generated from 686,000 isolated molecules. Although the general operation of the \gls{HRAFM} is common to all instruments, operational parameter settings (cantilever oscillation amplitude, tip--sample distance, CO tilt stiffness) lead to variations in the contrast observed on the resulting images. The value of the first two can be adjusted by modifying the microscope settings to enhance different features of the image. However, the latter depends on the nature of the tip, i.e. the differences in the attachment of the CO molecule to the metal tip that have been consistently observed and characterised in experiments~\cite{liebig2020quantifying,weymouth2014quantifying}. In order to cover the widest range of variants in the \gls{AFM} images, six different values for the cantilever oscillation amplitude, four for the tilt stiffness of the CO molecule and 10 tip-sample distances were used to generate \gls{AFMD}, resulting in a total of 240 simulations from each structure. 
We use the stack of 10 images resulting from the different tip--sample distances in a single input and the 24 parameter combinations as a data augmentation technique. That is, we feed the network with different image stacks randomly selected from the combinations of simulation parameters in each of the epochs for each of the molecules.

\subsection{IUPAC Tokenization} \label{Subsec:IUPAC_DATA}

Deep learning has already proven to have an extraordinary capacity to analyse data. This capacity is such that, in many cases, the biggest problem to be solved lies in defining an appropriate descriptor rather than in improving the existing analysis capacity. 
This is the case for \gls{AFM} images, where the complexity to design the output of a model is due to the existence of infinite molecular structures. To establish a model output that is unambiguous, uniform and consistent for the terminology of chemical compounds, we have adopted the \gls{IUPAC} nomenclature. 
Then, we have turned the standard classification problem~\cite{Carracedo2021MDPI} for a finite number of molecular structures into an image captioning task, developing a model that manages to formulate the \gls{IUPAC} name of each molecule.

\begin{figure}[t]
\centering
\includegraphics[clip=true, width=1.0\columnwidth]{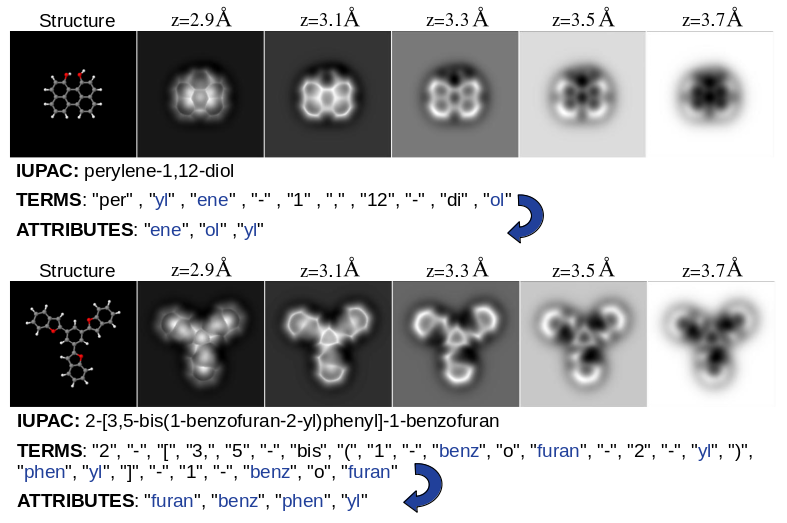}
  \caption{Two molecular structures with five of their associated AFM images at different tip-sample distances, the IUPAC name, the term decomposition of that name, and the associated attributes (a selection of 100 IUPAC terms that represent common functional groups, or chemical moieties, see text). The top structure shows that the attributes are sorted by length and alphabetically, not by the position in which they appear in the term decomposition. The bottom structure shows that attributes appear once even if they are repeated in the term decomposition.}
\label{Fig:IUPAC_LSTM_OUTPUT}
\end{figure}

Most image captioning techniques to describe images through language consist of a loop that predicts a new word at each iteration (time step). Our goal is to transfer this idea to the identification of AFM images through the \gls{IUPAC} formulation. Therefore, instead of predicting words at each time step, our model has to predict segments of the molecule's \gls{IUPAC} name (see~\cref{Fig:IUPAC_LSTM_OUTPUT}). That is, the set of tokens used to decompose each name are sets of letters, numbers and symbols that we call \textit{terms} and are used by the \gls{IUPAC} nomenclature to denote functional groups, to assemble additive names or to specify connections. Different combinations of these terms generate IUPAC names for the molecules, as exemplified in~\cref{Fig:IUPAC_LSTM_OUTPUT}.

A systematic split of the \gls{IUPAC} names in \gls{AFMD} reveals that some of the terms have a very small representation, not  enough to train a \gls{NN}.  We have discarded those that are repeated less than 100 times in \gls{AFMD}, retaining a total of 199 terms (see table~S1).  Consequently, we have also removed the molecules that have any of these terms in their \gls{IUPAC} name. 
In addition, we have dropped the molecules whose term decomposition has a length longer than 57, as there is not enough representation of such names in \gls{AFMD}. Even so, the set of annotations still contains 678,000 molecules, that we have split into training, validation and test subsets with 620,000, 24,000 and 34,000 structures, respectively. 

Our first attempts based on feeding a single model with a stack of AFM images provides poor results predicting the \gls{IUPAC} nomenclature. For this reason, we decompose the problem into two parts and assign each objective to a different \gls{NN} (see \cref{Fig:IUPAC_LSTM_OUTPUT,Fig:FLOW_M-ARNN} and section~\ref{Sec:MODEL_IUPAC} for a detailed description). We define the \textit{attributes} as a 100-element subset of the \gls{IUPAC} terms (see table S1) which mainly describes the most common functional groups in organic chemistry and, thus, are repeated a minimum number of times. The first network, named \gls{M-RNN-AT}, uses as input the stack of \gls{AFM} images and its aim is to extract the attributes, predicting  the main functional groups of the molecule (see \cref{Fig:IUPAC_LSTM_OUTPUT,Fig:FLOW_M-ARNN}). The second network, named \gls{AMRNN}, takes as inputs both the \gls{AFM} image stack and the attribute list with the aim of ordering them and complete the whole \gls{IUPAC} name of the molecule with the remaining terms which are not considered attributes (see \cref{Fig:FLOW_M-ARNN}B).
%

\gls{M-RNN-AT} reports information neither on the order nor the number of times that the {\em attribute} appears in the formulation. However, this first prediction plays a key role in the performance of the model. Unlike most of the \gls{NLP} challenges, the \gls{IUPAC} name completely identifies the structure and composition of the molecule. Thus, a prior identification of the main functional groups, not only releases the \gls{CNN} component of the \gls{AMRNN} from the goal of identifying these moieties, but, more importantly, almost halves the number of possible predictions of the \gls{AMRNN}. By feeding the \gls{AMRNN} with the attributes that are present in the \gls{IUPAC} name (predicted by the \gls{M-RNN-AT}), we are also effectively excluding the large number of them that do not form part of it. This is an extremely simple relationship that the network learns and that improves significantly its performance.

\subsection{Multimodal and Attribute Multimodal Recurrent Neural Networks (M-RNN$_\text{A}$ and AM--RNN )} \label{Sec:MODEL_IUPAC}


The standard approach for image captioning is based on an architecture that integrates a \gls{CNN} and a \gls{RNN}~\cite{mao2014deep,vinyals2015show}. Here, we focus on the well--known \gls{M-RNN}, which integrates three components (see fig.~S1). The \gls{CNN} encodes the input image into a high--level feature vector whereas the \gls{RNN} component has two key objectives: Firstly, to embed a representation of each word based on its semantic meaning and, secondly, to store the semantic temporal context in the recurrent layers. The remaining component is the multimodal $(\varphi)$  component, which is in charge of processing both \gls{CNN} and \gls{RNN} outputs and generating the output of the model.

\begin{figure}[b!]
\centering
\includegraphics[clip=true, width=1.03\columnwidth]{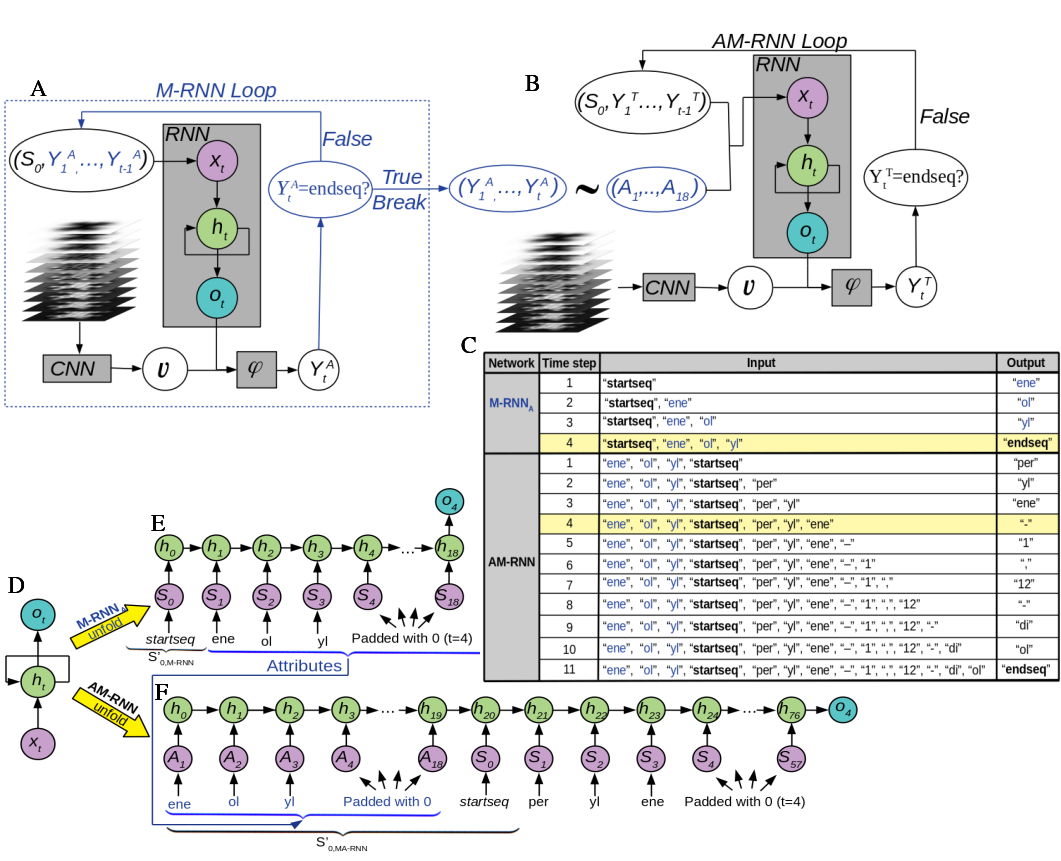}
 \caption{The architecture proposed for molecular identification through the IUPAC name is a composition of two networks (\gls{M-RNN-AT} and \gls{AMRNN}) whose data flow is shown in panels A and B. The gray square boxes represent each component of the models:  a convolutional neural network CNN, a recurrent neural network RNN, and the multimodal component $\varphi$. The arrows indicate the data flow within the model. \gls{M-RNN-AT} predicts an attribute at each time step until the loop is broken with the \textit{endseq} token (blue-line printed), whereas the \gls{AMRNN} predicts the sorted terms that give rise to the \gls{IUPAC} name.
 Panel C shows the inputs and outputs at each time step predicted by the  \gls{M-RNN-AT} and \gls{AMRNN} networks from a 3D image stack corresponding to the perylene-1,12-diol molecule. 
Panels D, E and F show a representation of the \gls{RNN}, in the same format used panels A and B, corresponding to the  fourth time step in \gls{M-RNN-AT}  and  \gls{AMRNN} for the  perylene-1,12-diol molecule. This figure highlights the fact that the state of the \gls{RNN}, in particular, the recurrent layer, depends  on the previous predictions.}
\label{Fig:FLOW_M-ARNN}
\end{figure}

As discussed in section~\ref{Subsec:IUPAC_DATA}, we have developed an architecture composed of two \gls{M-RNN}s (see \cref{Fig:FLOW_M-ARNN}A and \cref{Fig:FLOW_M-ARNN}B). 
The first one, the \gls{M-RNN-AT}, predicts the \textit{attributes} that are incorporated as input to the second one, the \gls{AMRNN}, which performs the \gls{IUPAC} name prediction. Although both \gls{AMRNN} and \gls{M-RNN-AT} are based on the standard \gls{M-RNN}~\cite{mao2014deep}, we introduce substantial modifications in each component. In \cref{Fig:FLOW_M-ARNN}A and \ref{Fig:FLOW_M-ARNN}B, we show the inputs for each component. The input of the \gls{CNN} component is a stack of 10 \gls{AFM} images, whereas the input of the multimodal component $\varphi$ consists of a concatenation of the outputs of the \gls{CNN} and \gls{RNN} components. 
To explicitly define the inputs of the \gls{M-RNN} components, it is worth recalling that a \gls{M-RNN} processes time series, so it will perform a prediction (\textit{attribute} or \textit{term}) at each time step. Let us start by defining the inputs of the \gls{RNN} component of the \gls{M-RNN-AT}.
We encode the \textit{attributes} of the model by assigning integer numbers (from 1 to 100) to each {\em attribute}.
The input of \gls{RNN} is a vector of fixed size 19, the maximum number of different attributes in the names of the molecules in \gls{AFMD} (17) plus the \textit{startseq} and  \textit{endseq} tokens. In the first step, it will contain $S_{0,\text{M-RNN$_{A}$}}=\textit{startseq}$ 
to provide the model with the information that a new prediction starts. This input is padded with zeros until we obtain a length of 19  (see figs.~\ref{Fig:FLOW_M-ARNN}A, \ref{Fig:FLOW_M-ARNN}C and \ref{Fig:FLOW_M-ARNN}E)  
and then processed by the \gls{RNN} component while the stack of \gls{AFM} images are processed by the \gls{CNN} component, each of them encoding the respective input into a vector. The two resulting vectors are used to feed the multimodal component $\varphi$, where they are concatenated and processed in a series of fully connected layers to finally produce a vector of probabilities 
(see Figs.~S1 and S2 for details on the \gls{RNN} and $\varphi$ layers). In this way the prediction at each time step corresponds to the most likely {\em attribute} $Y^A_1$ which replaces the padding zero of the corresponding position in the input sequence of the \gls{RNN} component in the next time step. 
This process is repeated until the {\em endseq} token is predicted, which breaks the loop. That is, for a given time step $t$, we feed the \gls{RNN} component of \gls{M-RNN-AT} with the input  $(S_0,Y^A_1,...,Y^A_{t-1})$, that concatenates the \textit{starseq} token $S_0$ with all the predictions already performed in previous time steps, and is padded with zeros until we obtain a length of 19 (see \cref{Fig:FLOW_M-ARNN}E for the example of $t=4$ in the indentification of perylene-1,12-diol molecule). 
Once the model has already predicted the $N_A$ \textit{attributes} it has to break the loop, so its last prediction must be the \textit{endseq} token (see \cref{Fig:FLOW_M-ARNN}C).

Once the prediction of the {\em attributes} has finished, the \gls{AMRNN} starts to operate in order to predict the \gls{IUPAC} name of the molecule.
For the input of the \gls{RNN} component and the prediction flow, we follow the same reasoning applied to \gls{M-RNN-AT}, replacing $S_{0,\text{M-RNN$_{A}$}}$ by $S_{0,\text{AM-RNN}}=(Y^A_1,..,Y^A_{18},\textit{startseq})$ (\cref{Fig:FLOW_M-ARNN}B).  
Each \gls{RNN} input is a vector of 76 components, arising from the concatenation of 18 \textit{attributes}  $(Y^A_1,..,Y^A_{18})\sim (A_1,...,A_{18})$ (padded if necessary) with the \textit{startseq} token and the predictions performed at each previous time step, $(Y_1,...,Y_{t-1})$, padded until we obtain a vector with length 57 --the maximum number of terms in the decomposition of the \gls{IUPAC} names in \gls{AFMD}-- (see Figs.~\ref{Fig:FLOW_M-ARNN}C and F).
Similarly as in the \gls{M-RNN-AT}, the semantic input is processed by the \gls{RNN} component while the \gls{AFM} image stack is processed by the \gls{CNN}, encoding the respective input into a vector. 
 The multimodal component $\varphi$ processes the \gls{CNN} output $v$,  concatenates the result with the output of the \gls{RNN}, and process this combined result producing a vector of probabilities as output of the network. 
 (see Figs.~S1 and S2 for details). The position of the larger component in the vector provides us with the prediction of the new {\em term} $Y_t$. The process stops when the \textit{endseq} token is predicted (see~\cref{Fig:FLOW_M-ARNN}C).

Denoting both $S_{0,\text{\gls{M-RNN}}}$ and $S_{0,\text{\gls{AMRNN}}}$  by $S_0'$, the data flow of both \gls{M-RNN} and \gls{AMRNN} is described by the following recurrence rules:
\begin{equation}
\begin{split}
x_1 & = S_0' \\
v & = \text{CNN}(I) \\
h_t & = \text{RNN}(h_{t-1},x_t)\\
Y_t & = \varphi(v,h_t), \ t\geq 1\\
x_t & = (S_0', Y_1,...,Y_{t-1}), \ t >1\\
\end{split}
\end{equation}

A more detailed description of each layer and the training strategy, far from trivial when combining a \gls{CNN} and a \gls{RNN},  can be found in sections S2 and S3, respectively.




\section{Results} \label{Sec:Results}


\begin{figure}[pt]
\centering
\includegraphics[clip=true, width=1.0\columnwidth]{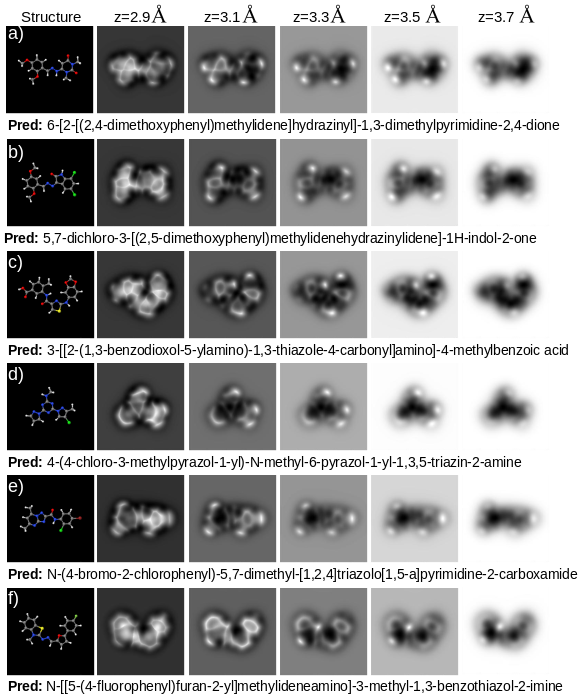}
 \caption{Set of perfect predictions (4-gram scores of 1.00). Each subfigure shows the molecular structure on the left, the \gls{AFM} images at various tip--sample distances on the right and the prediction, which matches exactly the ground truth, below the images.}
\label{Fig:Correct_iupac_pred}
\end{figure}

We have benchmarked the model by testing the trained networks with the 34,000 molecule test set, corresponding with 816,000 inputs from \gls{AFMD} associated to variations of the simulation parameters. 
The predicted \gls{IUPAC} names are identical to the annotations for $43\%$ of the molecules. 
Taken into account the complexity of the problem, we can consider this as a good result. 
Notice that each matching means that the model has identified from the images, without any error, all the molecular moieties and it has also provided the exact \gls{IUPAC} name, character by character, as shown in \cref{Fig:Correct_iupac_pred}. Our model is able to identify planar hydrocarbons, both cyclic or aliphatic, but also more complex structures as those including nitrogen or oxygen atoms that, due to their fast charge density decay~\cite{ellner2019molecular}, usually appear on the images as faint features (see for example \cref{Fig:Correct_iupac_pred}). Halogens, characterized on the images by oval features whose size and intensity are proportional to their $\sigma$--hole strength~\cite{Tschakert2020natcomm}, can be also correctly labeled (figs.~\ref{Fig:Correct_iupac_pred}b, \ref{Fig:Correct_iupac_pred}d and \ref{Fig:Correct_iupac_pred}e). The model can even recognize the presence of the fluorine element, that does not induce a $\sigma$--hole and, when bonded to a carbon atom, produces an \gls{AFM} fingerprint that is very similar to the one of a carbonyl group (compare~\cref{Fig:Correct_iupac_pred}e with \cref{Fig:Correct_iupac_pred}f). More surprisingly, hydrogen positions are often guessed, what is striking since hydrogen atoms bonded to $sp^2$ carbon atoms are hardly detected by the \gls{HRAFM} due to their negligible charge density~\cite{zahl2021TMA,shimizu2020effect}. 
Thus, many kinds of molecules, over half of our test set, including those showing non--trivial behaviors, have been correctly recognized by our model. 

\begin{figure}[pt]
\centering
\includegraphics[clip=true, width=0.9\columnwidth]{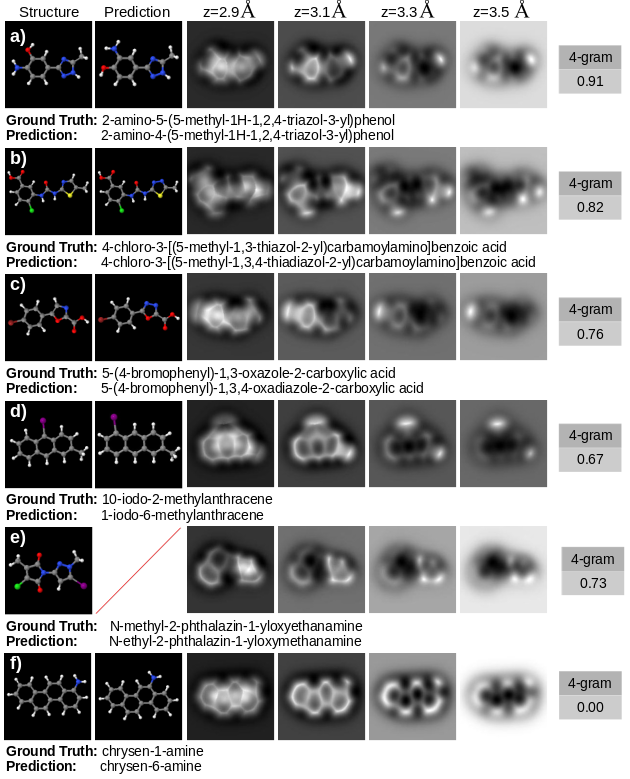}
 \caption{Examples of incorrect predictions, reflecting how the evaluation algorithm penalises errors. Each subfigure shows, from left to right, the simulated structure, the structure that matches the model prediction (where it exists), a set of \gls{AFM} images at various tip--sample distances and the 4--gram score. Below each subfigure is the \gls{IUPAC} name of the molecule (ground truth) and the prediction performed by the model (prediction).}
\label{Fig:Error_iupac_pred}
\end{figure}

However, this statistic does not reflect the real accuracy of the model. A deeper analysis of the results shows that its quality and usefulness is much higher than the naked figure of 43$\%$ could indicate.
Figure~\ref{Fig:Error_iupac_pred} shows that, even in those cases where the prediction is not correct, the majority of the examples still provide valuable information about the molecule. In order to quantify the accuracy of the prediction, we apply the n-grams of \gls{BLEU}~\cite{papineni2002bleu} (see~\cref{Fig:Error_iupac_pred}). This method, commonly used for assessing accuracy in \gls{NLP} problems, 
calculates the accuracy based on \textit{n--grams} of \textit{terms} between predicted and reference sequences. An \textit{n--gram} scores each prediction by comparing the sorted n--word groups appearing in the prediction with respect to the references. In our scenario, the comparison is with one single reference (ground truth), so it compares the common groups of \textit{n terms} that appear in both the prediction and the reference (for example,  perylene-1,12-diol, 4--gram reference groups include: ``per, yl, ene, -'', ``yl, ene, -, 1'',``ene, -, 1, ,'', etc).

First, we assess the accuracy of the \gls{M-RNN-AT}, the one that predicts the \textit{attributes}, i.e., the molecular moieties. A perfect match on the 1--gram's means that every \textit{attribute} in the reference appears in the prediction and that the prediction does not contain any other \textit{attributes}. 
Our model scores a 0.95 under this assumption. This is a very high mark that means this network does recognize the molecular components on 95\% of the cases. This result answers one of the more challenging open questions in the field~\cite{GrossAC2018,HeijdenACSNano2016,ellner2019molecular}: it demonstrates that the 3D \gls{HRAFM} data obtained with CO terminated apexes carries information of the chemical species present on the molecules, at least on the simulated images sets. 


\begin{table}[t!]
\centering
\begin{tabular}{| c| c| c | c | c |}
\hline
\hline
 Metric & 1--gram & 2--gram & 3--gram & 4--gram \\ 
\hline
\hline
Score & 0.88 & 0.84 & 0.79 & 0.76 \\  
\hline
\end{tabular}
\caption{\gls{BLEU} cumulative n--gram scores obtained with \gls{AMRNN} applied to \gls{AFMD}. The test has been performed on the set of 34,000 structures and their 24 combinations of simulation parameters.}
\label{Table:BLEU_n-grams}
\end{table}

For the assessment of the overall prediction of the model, we propose the cumulative 4-gram, a common metric for the evaluation of linguistic predictions.
This metric weights the scores obtained with the 1,2,3,4--grams and also performs a product with a function that penalises the different lengths between prediction and reference. \gls{BLEU} scores (see \cref{Table:BLEU_n-grams}) reveal that \gls{AMRNN} also performs exceptionally well. Note that, in this case, the 1--gram shows the set of \textit{terms} that are in both prediction and reference. That is, despite not providing the correct formulation, the model is able to predict 88\% of the \textit{terms} that the name contains, in agreement with the prediction capability showed by our first \gls{M-RNN-AT}, and indicating a great deal of chemical information about the molecule. In addition, \gls{AMRNN} scores 0.76 in the evaluation with the cumulative 4-gram, assessing large segments of the \gls{IUPAC} name. Fig.~\ref{Fig:Error_iupac_pred} puts the accuracy of the model based on this assessment in context with a set of examples with different scores. 
Note that  
\cref{Fig:Error_iupac_pred}f shows a frequently occurring case where, by applying a metric developed to assess translation in longer texts with several references, mistakes in predictions composed of only a few \textit{terms} are overly penalised.
\Cref{Table:BLEU_4_gram_decomposition} provides a systematic study of this limitation of the metric, showing an analysis of the score obtained by splitting the test set according to the number of terms into which the corresponding IUPAC name decomposes. The accuracy of the model is worse in molecules whose term decomposition is shorter. The reason for this seemingly contradictory fact is that the cumulative 4-gram metric penalises more for errors in short chains. 
As shorter strings contain fewer subgroups of 4 terms, the 4--gram scoring method penalises an error in a smaller chain more heavily than in a longer one (as shown in \cref{Fig:Error_iupac_pred}a and \cref{Fig:Error_iupac_pred}f). 

Comparing the predictions with the references on a term--by--term basis, we find that the 25.1\% of the errors are due to misclassification of one number term with another number, i.e. misplacing a group of atoms, and the 17.1\%, 4.8\%, 4.7\% and 2.7\% of the errors are due to a misclassification of the ``-'', ``('' or ``)'', ``yl'' and ``['' or ``]'' \textit{terms}, respectively. Therefore, almost half of the errors made are located in the prediction of characters more related to the chemical formulation than to the information extracted from the images. Moreover, we must point out the fact that when the model predicts incorrectly, it sometimes generates \gls{IUPAC} names that do not correspond to any molecule (see \cref{Fig:Error_iupac_pred}e).  
These results indicate that it is not the capability of our model to recognise the molecules but the ability of the \gls{RNN} component to properly write the name what is limiting the success rate.  This conclusion is consistent with a recent work where automated \gls{IUPAC} name translation from the SMILES nomenclature~\cite{SMILES}, that completely characterizes the structure and composition of a molecule, is done by a \gls{RNN}~\cite{rajan2021stout}, obtaining just a \gls{BLEU} 4--gram score of 0.86.

Deep learning architectures are developed based on human intuition to improve the accuracy of the model. However, it is difficult to analyse in detail why the model succeeds or fails. When representing the input \textit{terms} in the \gls{RNN} component, we apply a word embedding that is trained with the rest of the model (see figs.~S2 and S3). Previous research has shown that representations in this space capture the semantic meaning of words and establish algebraic relationships between them~\cite{pennington2014glove,chen2020simple, minaee2021deep}. 
It is truly remarkable to see that these results have been transferred to the formulation, grouping the \textit{terms} according to their semantic meaning or according to the interactions described by the image stacks. 
We have verified this by projecting each of the \textit{terms} into the 32--dimensional embedding space belonging to the \gls{AMRNN}, defining a L2 norm and computing the distances between the \textit{terms}. These results show that \textit{terms} with similar semantic meaning are close together (see fig.~S9). For example, the closest \textit{terms} to brom are chlor, fluor, and iod, or the \textit{terms} closest to nona are octa, deca, undeca and dodeca. 
This also reflects in the fact that the \textit{terms} that the model most commonly gets wrong are the closest ones, such as the errors in the prediction of the numbers that place atomic groups in specific positions (see figs.~\ref{Fig:Error_iupac_pred}a, \ref{Fig:Error_iupac_pred}d and \ref{Fig:Error_iupac_pred}f), or the mistaken of one halogen for another. In other words, the erroneous terms have, in general, a similar semantic meaning.


\begin{table}[t!]
\begin{center}
	\begin{tabular}{| c| c| c | c | c | c | c |}
\multicolumn{7}{ c }{\bfseries Number of \textit{terms}}  \\ \hline	
Term Decomposition & 0--10 & 10--20 &20--30 &30--40 &40--50 &50--60 \\ \hline \hline
4--gram score & 0.59 & 0.73 & 0.78 & 0.79 & 0.75 & 0.66 \\ \hline
\multicolumn{3}{ c }{ } \\
\end{tabular}
\begin{tabular}{| c| c| c | c | c |}
\multicolumn{5}{ c }{\bfseries Atom height difference}  \\ \hline	
Distance & $<$0.5 \r{A} &  $<$1.0 \r{A} &  $<$1.5 \r{A} &  $\geq$1.5 \r{A} \\ \hline \hline
4--gram score & 0.79 & 0.62 & 0.62 & 0.50\\ \hline 
\end{tabular}
\captionof{table}{Score with the \gls{BLEU} cumulative 4--gram metric based on the characteristics of the molecules and their annotations.  The top table divides the scores into subsets based on the length of the string of \textit{terms} into which the \gls{IUPAC} name is broken down. The bottom table divides the test set score into subsets based on the maximum difference in heights between atoms in the molecule (excluding hydrogens). }
\label{Table:BLEU_4_gram_decomposition}
\end{center}
\end{table}

Non--planar structures are a challenge for AFM-based molecular identification. \Cref{Table:BLEU_4_gram_decomposition} shows an analysis of the score obtained by splitting the test set according to  different ranges of molecular torsion. 
In line with the conclusions reached in ref.~\cite{alldritt2020automated}, our model has a hard work to fully reveal the structure of molecules whose height difference between atoms exceeds 1.5 \r{A}. This is an expected result as the microscope is highly sensitive to small variations on the probe--sample separation, and the interaction becomes highly repulsive on a distance range of 50--100 pm, inducing large CO tilting and image distortions. This makes very difficult to get a proper signal from lower atoms on molecules with non--planar configurations. However, most of these molecules would adopt a flatter configuration upon surface adsorption. We have tested our model by randomly selecting four of the non--planar structures whose prediction scores an arithmetic mean of 0.40 in the cumulative prediction of the 4--gram. We force them to acquire a flat structure and, then, we run the test again (see figs. S6 and S7).
Prediction scores improve in a range of 0.2 to 0.55, resulting in a new mean cumulative 4--gram of 0.73.  This improvement  represents semantically going from a prediction that barely provides any useful information about the molecule to one that in many cases gets it absolutely right. 
Thus, while the model already scores very high in the test with simulated images of gas--phase molecules, the performance would definitely improve with the flatter configurations expected for the adsorbed molecules measured in the experimental \gls{HRAFM} images.

\section{Discussion} \label{Sec:Discussion}

The results presented so far show that the stacks of 3D frequency shift images contain information not only on the structure of the molecules but also regarding their chemical composition. This information can be extracted by deep learning techniques, which, additionally, are able to provide the \gls{IUPAC} name of the imaged molecules with a high success rate. 
Our combination of two M--RNNs is able to correctly recognize the molecule in many cases, even in those where it is difficult to discern between similar functional groups --as fluorine terminations with either carbonyl or even -H terminations--, or in image stacks where some moieties provide very subtle signals (see \cref{Fig:Correct_iupac_pred}). 
Some mistakes  do appear from the chemical recognition point of view, especially in those molecules showing significant non-planar configurations where the performance is lower (see figs.~S6 and~S7 together with \cref{Table:BLEU_4_gram_decomposition}). However, apart from these fundamental drawbacks, most of the errors in the predictions are related to the spelling of \gls{IUPAC} names. That is, misplacement of functional groups or the incorrect use of parentheses, square brackets or hyphen characters, etc. It seems that these errors are frequent for RNNs dealing with the \gls{IUPAC} nomenclature~\cite{rajan2021stout}. 

At this stage, it is worth considering if other DL architectures or alternative chemical nomenclatures could improve the molecular identification based on \gls{HRAFM} images.
We have already pointed out that, leaving out the additional problem of extracting the chemical information from the images, a \gls{RNN} only achieves a \gls{BLEU} 4--gram score of 0.86 when translating from the SMILES to the IUPAC name~\cite{rajan2021stout}. Nomenclature translation has been addressed with architectures based on the novel transformer networks~\cite{attention2017}, obtaining a practically perfect accuracy~\cite{krasnov2021struct2iupac, handsel2021translating}. Also, automatic recognition of molecular graphical depictions is able to correctly translate them to their SMILES representation with a 88\% or 96\% accuracy by using either a standard encoder-decoder~\cite{Img2MOL} or a transformer~\cite{rajan2021decimerv1} network. 
However, in our work we face a different problem since we deal with identification from \gls{AFM} images instead of either molecular depictions, that contains all the chemical information needed to name a molecule, or translation between nomenclatures. Furthermore, the application of transformers to the identification from \gls{AFM} images is not straightforward. First, tokenization must be consistent and each term must have a chemical meaning so that the embedding layers learn a meaningful information representation (see Fig. S9). This point has only been considered in ref.~\cite{krasnov2021struct2iupac}. More importantly, our method achieves high accuracy due to the initial attribute detection, forcing us to develop an architecture that is, in principle, incompatible with transformers, which are based on encoder-decoder networks.

Regarding other nomenclature systems for describing organic molecules, besides the already cited SMILES~\cite{SMILES}, there are other proposals such as  InChI~\cite{INCHI} or SELFIES~\cite{Krenn_2020}, 
whose textual identifiers use the name of the atoms and bond connectivity. These systems miss relevant chemical information that is not provided by describing the molecule as a set of individual atoms rather than as moieties made up of atoms. 
Unlike these systems, the \gls{IUPAC} nomenclature is focused on the classification of functional groups, an approach consistent with the characteristics shown by the \gls{AFM} image features, that reflects in our proposal of a dual architecture composed by \gls{M-RNN-AT} and \gls{AMRNN}. 
The SELFIES nomenclature establishes a robust representation of graphs with semantic constraints, solving some problems that arise in computer writing with other nomenclatures. However, the atom-based description would force an approach without attribute prediction, which is the key to obtain a high accuracy with our model.
Hence, it seems to be a trade-off between the limitations and improvements offered by these nomenclatures, suggesting that a dramatic improvement in performance is not expected, although further work is needed in order to reach a final conclusion.

Finally, we should point out that our analysis so far has been based on simulated \gls{HRAFM} images. 
In ref.~\cite{Carracedo2021MDPI}, we showed that the experimental images contain features that are not reflected in the theoretical simulations. Data augmentation has been applied during the training of our model (see sections~S3.2 and S3.4) to capture these effects. Although limited by the scarcity of experimental results suitable to apply our methodology, we have obtained very encouraging results. 
%
We have selected constant--height \gls{AFM} images of dibenzothiophene and 2-iodotriphenylene from refs.~\cite{EXP} and~\cite{martin2019bond}, corresponding to 10 different tip--sample distances, covering a height range of 100 pm for dibenzothiophene (identical to the one spanned by the 3D stacks of theoretical images used to train our model) and 72~pm for 2-iodotriphenylene (see section~S5 for details). 
Despite the strong noise in the images and the white lines crossing the images diagonally (see fig.~S8), the prediction of dibenzothiophene is perfect, scoring 1.00 on the 4--gram, whereas for 2-iodotriphenylene the model predicts ``2iodtriphenylene", missing a hyphen but  providing all the relevant chemical information. Despite the good results obtained with these two experimental image stacks, no significant conclusions can be extracted given the very small size of the test. A larger, systematic analysis with a proper experimental data is necessary to further address the accuracy of our model.

\section{Conclusions}


In this work, we have shown how deep learning models, trained with the simulated \gls{HRAFM} 3D image stacks for 678,000 molecules included in the \gls{AFMD} dataset,  are able to perform full chemical-structural identification of molecules. Motivated by the unfeasibility of defining a classification in the usual sense of \gls{AI}, we turned the problem into an image captioning problem. Thus, instead of aiming to have a model that knows every atomic structure, we endow it with the ability to formulate. As a result the model is not only able to identify images that have not been previously shown to it, but it is also able to predict the \gls{IUPAC} name of these unknown structures. 
We have devised a two--step procedure involving the combination of two  \gls{M-RNN}s. In a first step, the \gls{M-RNN-AT} identifies the attributes, the most relevant functional groups present in the molecule. This initial step is already of importance because the algorithm provides useful information about the chemical characteristics of the molecule. In a second step, the \gls{AMRNN}, whose inputs arise from the \gls{M-RNN-AT}, sorts the information of the functional groups, adds extra characters (connectors, position labels, other tags, etc.) and the remaining functional groups that are not part of the \textit{attributes} set. That is, the \gls{AMRNN} assigns the positions of the functional groups, completes the remaining terms and writes down the final \gls{IUPAC} name of the molecule.

We have tested the model on a set of 816,000 \gls{AFM} images belonging to 34,000 molecules that have not been shown before to the network. The bare results point out that in a 43\% of the cases the predictions are exactly the same with respect to the reference in \gls{AFMD}, character by character. To further asses the usefulness of the wrong predictions by the model, we apply the metrics defined by the \gls{BLEU} n-gram. This is a robust methodology used in assessing the accuracy of our model that weights the accuracy of a prediction against a reference. The accuracy of the \textit{attribute} prediction assessed with the 1-gram scores 0.95, whereas the overall accuracy of the model is determined with the cumulative 4-gram, scoring 0.76. 
This high value means that, even when the model does not achieve a perfect prediction, it provides valuable chemical insight, leading to a correct IUPAC name of a similar molecule in the vast majority of the cases.
Finally, it is worth noting that this model could be applied to \gls{AFM} images obtained in experiments. 
Due to the lack of a systematic and large set of experimental images, we do not have definitive conclusions yet, but we have obtained very encouraging results in two examples.

\section*{Acknowledgments}
We deeply thank Dr.~D.~Martin-Jimenez and Dr.~D.~Ebeling for providing us with the experimental images of 2-iodotriphenylnene used to test the model. We would like to acknowledge support from the Comunidad de Madrid Industrial Doctorate programme 2017 under reference number IND2017/IND-7793 and from Quasar Science Resources S.L.
P. Pou and R.P. acknowledge support from the Spanish Ministry of Science and Innovation, through project PID2020-115864RB-I00  and the ``Mar\'{\i}a de Maeztu'' Programme for Units of Excellence in R\&D (CEX2018-000805-M). Computer time provided by the \gls{RES} at the Finisterrae II Supercomputer is also acknowledged.

\bibliographystyle{Science}
\bibliography{DL_AFM}
\addcontentsline{toc}{chapter}{Bibliography}

\end{document}



\baselineskip24pt

\maketitle

\newpage
\tableofcontents

\newpage


\section{IUPAC tokenization} \label{Subsec:IUPAC_DATA}

\begin{table}[b!]
\centering
\begin{adjustbox}{max width=1.0\textwidth}
\begin{tabular}{ | c |c | c | c | c | c | c | c | c | c |}
\hline 
acen & acet & acid & acr & alde & alen & amate & amide & amido & amim \\ \hline
amine & amino & amo & ane & ani & anil & ano & anone & anthr & ate \\ \hline
ato & aza & aze & azi & azido & azin & azo & azol & benz & brom \\ \hline
but & carb & chlor & chr & cyclo & ene & eno & eth & fluor & form \\ \hline
furan & furo & hyde & hydr & ic & ida & ide & idin & idine & ido \\ \hline
imid & imidin & imine & imino & ind & ine & ino & iod & iso & itrile \\ \hline
ium & meth & mine & naphth & nitr & nium & nyl & oate & oic & ol \\ \hline
ole & oli & olin & oline & olo & om & one & oso & oxo & oxol \\ \hline
oxy & oxyl & oyl & phen & phth & prop & pter & purin & pyr & pyrrol \\ \hline
quin & sulf & thi & tri & urea & yl & ylid & yridin & zin & zine \\ \Xhline{5\arrayrulewidth}
dodeca & undeca & lambda & phosph & cinnam & xanth & porph & coron & deca & guan \\ \hline
hept & amic & pent & enal & octa & anal & nona & mido & nida & azet \\ \hline
tetr & ulen & anol & hypo & pine & ysen & anth & tere & acyl & yrin \\ \hline
inin & tris & pino & mid & hex & nia & bis & per & nio & pin \\ \hline
ite & rin & alo & en & di & 11 & yn & an & cy & de \\ \hline
15 & 10 & 13 & 14 & 12 & in & 18 & 21 & az & al \\ \hline
bi & et & ep & id & ox & il & or & 16 & 17 & on \\ \hline
( & 6 & ) & - & [ & a & ] & N & ' & 1 \\ \hline
7 & 4 & 5 & 3 & 9 & 2 & 8 & H &   & , \\ \hline
o & b & e & c & C & d & O & g & f  & \cellcolor{gray!95}  \\ \hline
\end{tabular}
\end{adjustbox}
\caption{Table of terms for \gls{IUPAC} decomposition. The elements above the bold line are the subset of 100 attributes considered. The grey cell does not correspond to any term, it has been coloured in order to distinguish it from the term that spells an empty space between two words.}
\label{Tab:Terms}
\end{table}


In order to tokenize the IUPAC nomenclature we selected a set of terms consisting of prefixes, suffixes, numbers, etc., whose combinations generates the 686,000 IUPAC names corresponding with the molecules belonging to \gls{AFMD}~\cite{QUAM-AFM_repository}. Deep learning models require large datasets in which each target is repeated many times to be learned during the training. We have analysed the number of times that each term appears in \gls{AFMD} and removed those terms that are repeated less than 100 times. This reduces the total number of terms considered to a total of 199, shown in \cref{Tab:Terms}. Consequently, we have discarded molecules whose decomposed IUPAC name contains any of these terms. 
Similarly, extremely long IUPAC names have very little representation in \gls{AFMD}, so we also discard molecules whose IUPAC name decomposes into more than 57 terms. In this way, we have slightly reduced the set of available inputs to a total of 678.000 molecules.

As explained in the main text, we combine two different  \gls{M-RNN}s to achieve molecular identification. The first network, named \gls{M-RNN-AT}, uses as input the stack of \gls{AFM} images and its aim is to extract the \textit{attributes}, a 100-element subset of the terms which are mainly used to designate functional groups and not positions (see \cref{Tab:Terms}). The second network,  \gls{AMRNN}, consists of an adaptation of the \gls{M-RNN} \cite{mao2014deep} for the \gls{IUPAC} nomenclature prediction, where we add an input of semantic attributes predicted by the \gls{M-RNN-AT} (see section~\ref{sec:MRNN_models} for details).
This strategy differs from the concept of {\em attention} \cite{bahdanau2014neural,parikh2016decomposable,raffel2017online},  implemented in other \gls{NLP} problems, and that has led to the development of the transformers \cite{attention2017,bert2018,GPT-32020,pan2020x}.
%
In this case we assume that the relationship between each IUPAC name and each molecule is biunivocal, (although there are some exceptions we suppose that the names of each compound have been entered under the same rules in Pubchem). This approach makes the problem posed different from most of the challenges faced in image captioning where there are multiple descriptions for the same image. Therefore, a prior identification of the main functional groups, not only releases the \gls{CNN} component of the \gls{AMRNN} from the task of identifying these moieties, but, more importantly, almost halves the number of possible predictions of the \gls{AMRNN}: By feeding the \gls{AMRNN} with the attributes that are present in the \gls{IUPAC} name (predicted by the \gls{M-RNN-AT}), we are also effectively excluding the large number of them that do not form part of it. This strategy improves significantly its performance.


\newpage
\section{M-RNN$_A$ and AM-RNN models: A layer description}\label{sec:MRNN_models}

\begin{figure}[b!]
\centering
\includegraphics[clip=true, width=1.0\columnwidth]{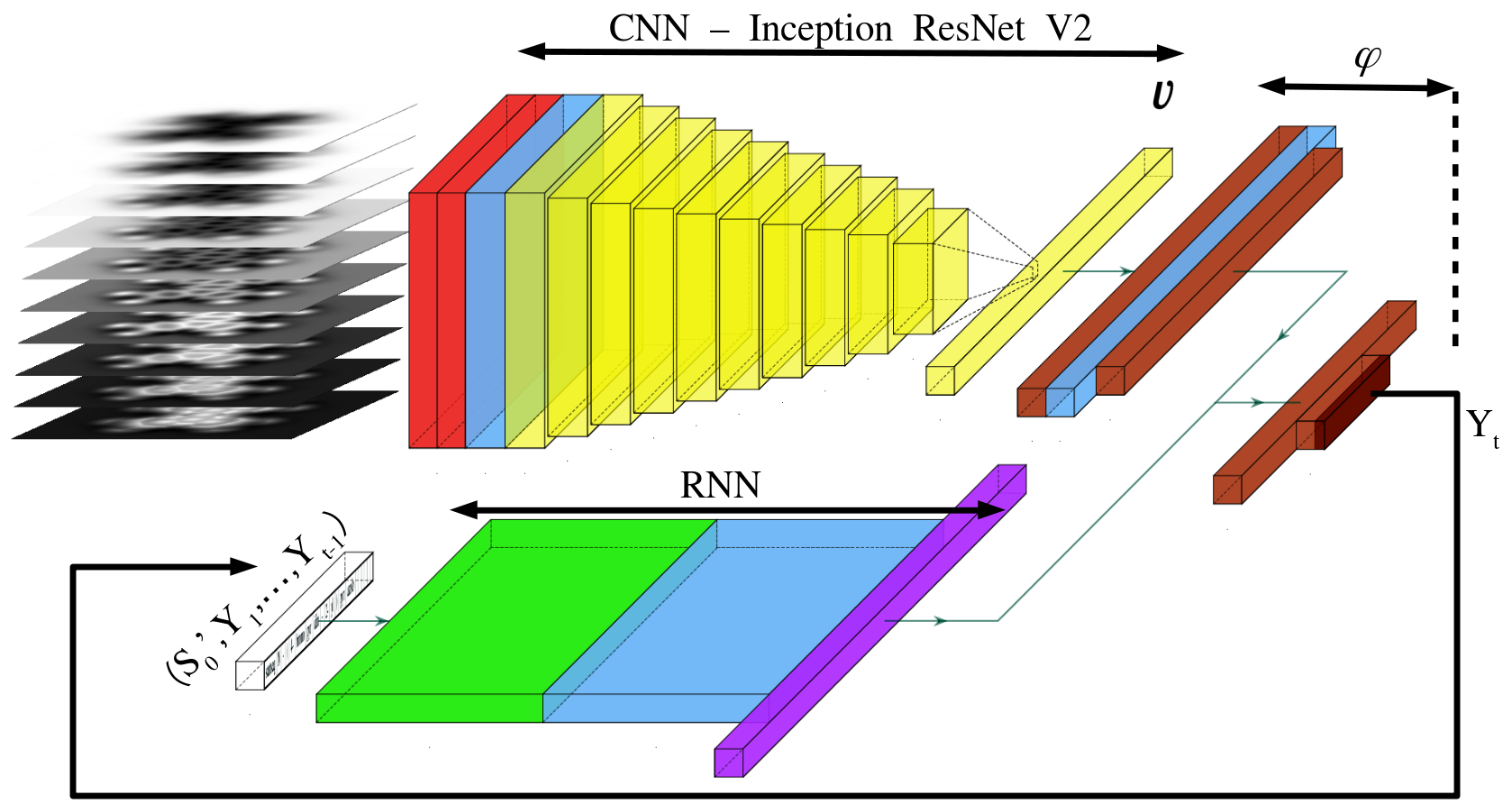}
\caption{Graphical layer representation of \gls{M-RNN-AT} and \gls{AMRNN} with the layers that constitute their three components, \gls{CNN}, RNN and $\varphi$. 
The CNN component follows the Inception ResNet V2 model, where the first 2D--convolutional layers have been replaced by two 3D convolutional layers (to process the image stack, shown in red), followed by a dropout layer (blue). The yellow blocks are just a pictorial representation of each block of the original Inception ResNet V2 model. Notice that the last fully connected layer of the model, which is specific for the original classification task, has been removed, obtaining an output vector ($v$) with length 1539.
The  RNN and $\varphi$ components include embedding (green), dropout (blue), fully connected (brown) and recurrent (purple) layers. The purple box represents a \gls{GRU} layer in \gls{M-RNN-AT}, whereas in the \gls{AMRNN} it represents a \gls{LSTM} layer. 
$S_0'$ represents the input for the RNN component in the first time step: the \textit{startseq} token in the \gls{M-RNN-AT} and the concatenation of the \textit{attributes} with the \textit{starseq} token in \gls{AMRNN}. The subsequent inputs include the {\em attributes} ({\em terms}) predicted in previous time steps by \gls{M-RNN-AT} (\gls{AMRNN}).
Although both \gls{M-RNN-AT} and \gls{AMRNN} have the same structure, each layer of RNN and $\varphi$ components has different dimensions (see \Cref{Fig:Model_details} for a detailed description).} 
\label{Fig:INC_RNN_AFM}
\end{figure}

\begin{figure}[b!]
\centering
\includegraphics[clip=true, width=0.8\columnwidth]{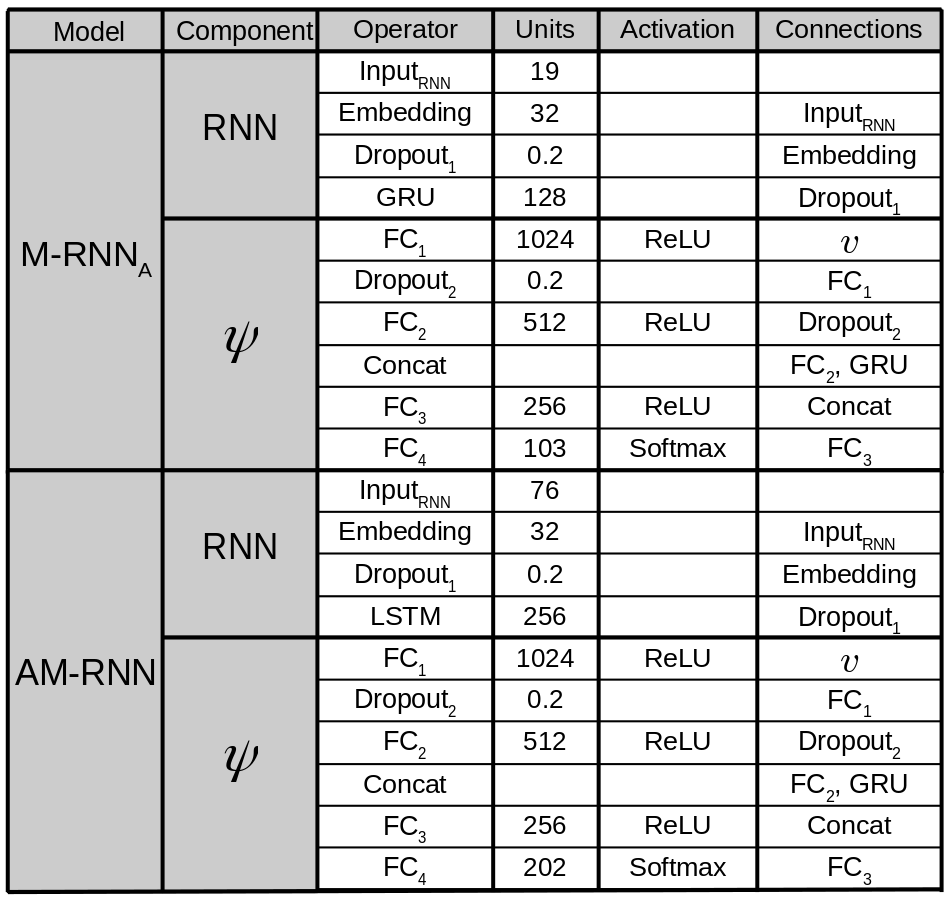}
\caption{Layer--by--layer details of the RNN and $\varphi$ components integrated in \gls{M-RNN-AT} and \gls{AMRNN}. $v$ denotes the output vector of the \gls{CNN} component. The layers are the same for both models, except for the recurrent one, a \gls{GRU} in \gls{M-RNN-AT} (attribute prediction) and an \gls{LSTM} in  \gls{AMRNN} (term prediction). }
\label{Fig:Model_details}
\end{figure}

As previously mentioned, the architecture developed for molecular identification with the \gls{IUPAC} nomenclature is composed of two models, \gls{M-RNN-AT} and \gls{AMRNN}, (see Figure 2 of the main text). Each one is composed by a \gls{CNN}, a RNN and a multimodal componet $\varphi$. 
\Cref{Fig:INC_RNN_AFM} displays  the type of layers that constitute each component of both \gls{M-RNN-AT} and \gls{AMRNN}. 
The architecture of the \gls{CNN} component is identical in both models, while RNN and $\varphi$ share the same structure and, except for the RNN layer, the same type of layers with different number of units (e.g. kernels in convolutional layers, vector length in fully connected layers, etc) according to the specific purpose of the model. 
\Cref{Fig:Model_details} provides the details, highlighting the differences between the layers of the RNN and $\varphi$  components of the two models.
%

The \gls{CNN} component consists of a modification of the Inception ResNet V2 model~\cite{szegedy2016inception}, identical for both \gls{M-RNN-AT} and \gls{AMRNN}. This well--known model has been developed to be applied to 2D images whereas in our case we process 3D maps (stacks of 10 \gls{AFM} images with various tip--sample distances). Therefore we have replaced the first 2D--convolutional layers in Inception ResNet V2 by two 3D convolutional layers, each one with 32 filters, (3,3,3) kernel size and (2,1,1) strides, followed by a dropout layer. We have verified that this dropout layer is essential for the model to generalize to different images, such as the experimental ones. In addition, we have removed the last fully connected layer of the model, which is specific for the original classification task, obtaining an output vector ($v$) with length 1539. 


The goal of the RNN component is to use sequential data to add new {\em terms} to the formulation. To this end, the architecture of this component is developed according to two key objectives: Firstly, to embed a representation of each {\em term} based on its semantic meaning and, secondly, to store the semantic temporal context in the recurrent layers. To perform the representation of each {\em term} in a vector space, the RNN component has an embedding layer that is able to capture relationships between {\em terms} (see \cref{sec:embedding} for a more detailed analysis). The embedding layer is followed by a dropout layer that acts as a regularizer and finally the data goes through a recurrent layer which, with the temporal context generated by all the predictions made in previous time steps, makes a proposal of predictions that is processed by the multimodal component (see \cref{Fig:INC_RNN_AFM}). The recurrent layer is a \gls{GRU} in \gls{M-RNN-AT} (attribute prediction) and an \gls{LSTM} in  \gls{AMRNN} (term prediction). 
The multimodal component $\varphi$ first processes the \gls{CNN} output $v$ in two fully connected layers with a dropout between them. Subsequently, this output is concatenated with the output of the RNN (see \cref{Fig:INC_RNN_AFM}). Finally, the result of the concatenation feeds two fully connected layers, the first one activated with \gls{relu} activation function whereas the second one is activated with \gls{softmax} activation function, that converts the outputs from the layer into a vector with components that represent probabilities that sum to one.


\newpage
\section{Model learning} \label{Sec:training_IUPAC}

Both \gls{M-RNN-AT} and \gls{AMRNN} could, in principle, be trained in an end--to--end process, however, this would require more than a year of training and it has been found that this type of training for M-RNN results in not providing detailed descriptions~\cite{gu2018stack}. 
To avoid it, we train each component of the models in different stages, fixing the weights of the \gls{CNN} and RNN alternatively while training the rest of the model.

In the first stage, both \gls{CNN} and RNN components are initialised with random weights, so if we fix the weights of the \gls{CNN} while training the RNN, the \gls{CNN} would perform a random representation of the input image stack, and consequently, the weights of the RNN component would be updated under random rules. An analogous reasoning can be applied for the reverse case, in which we would fix the weights of the RNN and train the \gls{CNN}. We solve this issue initialising the \gls{CNN} component with pre--trained weights. 
To determine this weights,  we perform a classification with the \gls{CNN} based on  classes of  molecules that shared the same chemical composition (number of different chemical species and number of atoms for each specie, excluding the H atoms), as described in detailed in~\cref{SubSec:Chem_Species_Data}.

\subsection{Molecular Classes for Transfer Learning} \label{SubSec:Chem_Species_Data}

\begin{figure}[h!]
\centering
\includegraphics[clip=true, width=1.0\columnwidth]{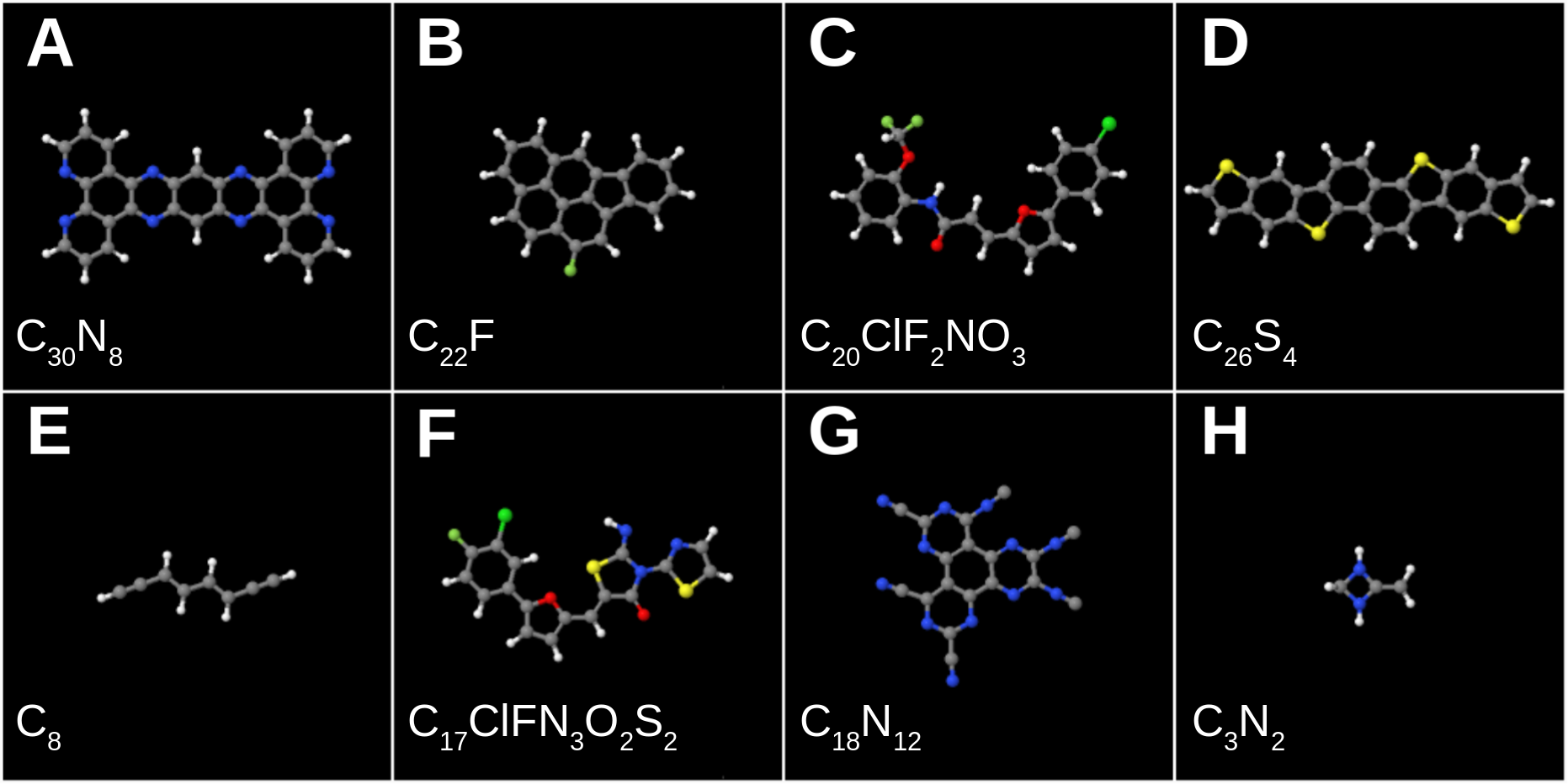}
 \caption{Atomic structures belonging to different classes for classification according to their chemical species.}
\label{Fig:Structures_JMOL}
\end{figure}

The classification of \gls{AFM} images defining the model output as each individual molecule is an impossible task because \gls{AFMD} does not include enough images of each particular structure and has an excessive number of molecules (classes). Thus, we simplify the problem by grouping molecules based on their chemical composition. Hence, we define the class of a molecule by the type of atomic species that it contains and the number of repeated atoms of each of these species. To obtain a representative number of images of each class, we exclude the hydrogens from the species list (see \cref{Fig:Structures_JMOL}), so that molecules with completely different structures such as pyrazine, pyridazine, but-2-enedinitrile or butanedinitrile belong to the same class ($C_4N_2$). 
This results in a total of 2339 classes for the molecule structures considered in \gls{AFMD}.

\subsection{Pre--training the CNN component} \label{Sec:Class_Inception_ResNet}

As mentioned above, it is necessary to pre-train the \gls{CNN} component with a classification that groups the molecules in  the \gls{AFMD}  dataset in classes describing its chemical composition, irrespective of their structure.
We perform this classification with the same Inception ResNet V2 model~\cite{szegedy2016inception} that we used for the \gls{CNN} element in the two \gls{M-RNN}s . To this end, we replace the first convolutional layer by two 3D convolutional layers followed by a dropout layer and, instead of removing the output layer as described in the main text for the \gls{NLP} target, we modify its number of units to 2339, corresponding to the number of classes defined in \cref{SubSec:Chem_Species_Data}.

\begin{figure}[t!]
\centering
\includegraphics[clip=true, width=1.0\columnwidth]{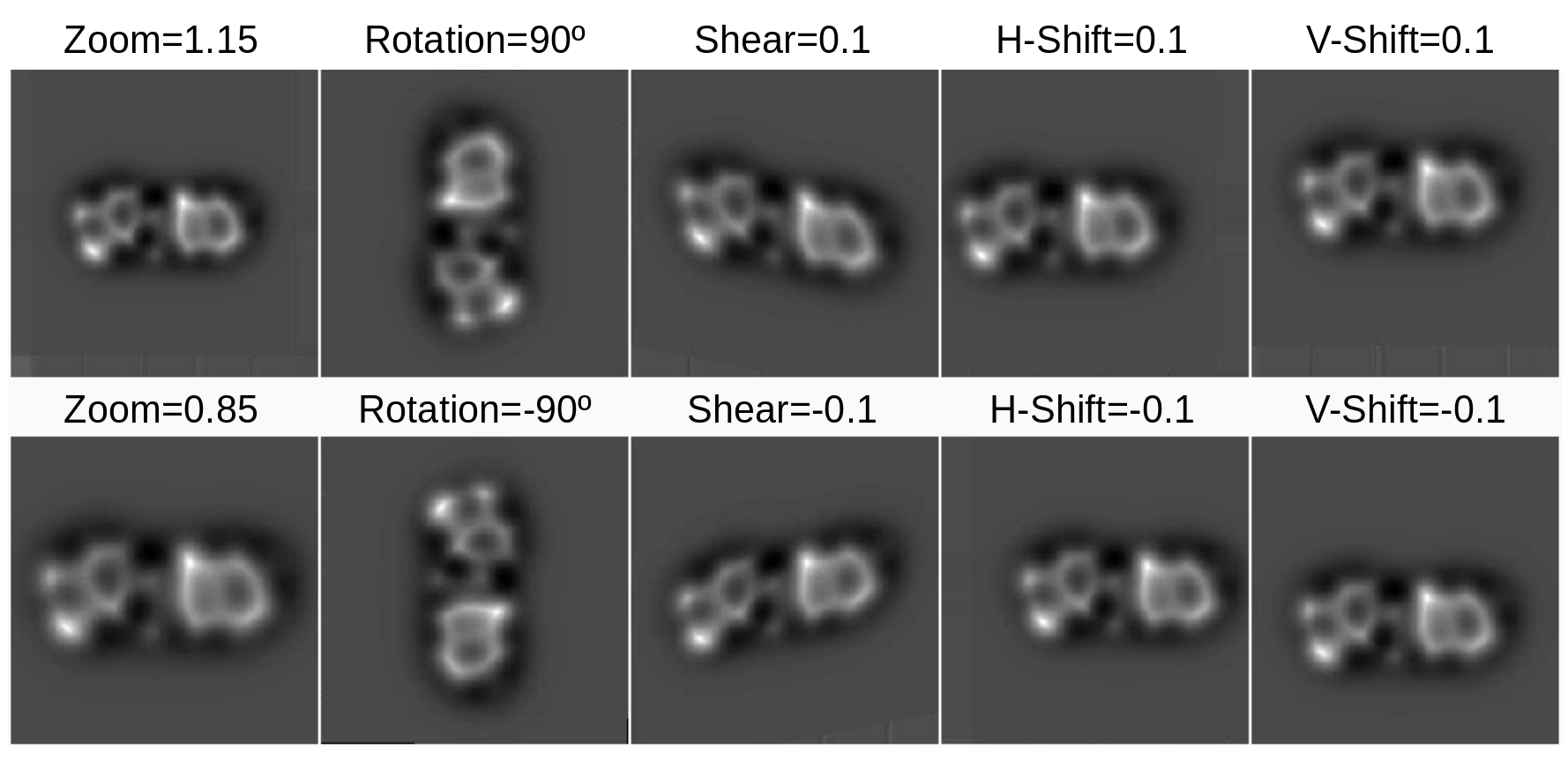}
 \caption{Results of applying the \gls{IDG} to the same \gls{AFM} image of a N-(5-amino-4-methylpyridin-2-yl)-6-fluoro-1-benzothiophene-2-carboxamide molecule.}
\label{Fig:Iupac_IDG}
\end{figure}

In each epoch, we select a single combination of simulation parameters from those included in the \gls{AFMD} dataset\cite{QUAM-AFM_repository} for each molecule. Therefore, the model receives inputs of the same molecules but different image stacks at each epoch. In addition, we apply an \gls{IDG} (see \cref{Fig:Iupac_IDG}) to the input image stack 
in order to simulate experimental features that are not captured in the theoretical simulations, as discussed in ref.~\cite{Carracedo2021MDPI}. 
The values selected for the distortions are randomly chosen in ranges of [1,360]--degree rotations, $\pm0.15$ zoom range,  $\pm0.1$ shear range, and $\pm0.1$ both vertical and horizontal shift range, as illustrated in \cref{Fig:Iupac_IDG}. When a molecule is rotated or moved during an \gls{AFM} experiment, all resulting images show similar variations, so we apply the same deformation parameters to the ten images that compose each stack. These variations, coupled with the dropout layer in the \gls{CNN} component, ensure that the model is robust regarding soft variations of the inputs that are present in experimental images.

We train the \gls{CNN} minimizing the error of the Negative Log Likelihood loss function with the \gls{Adam} optimizer~\cite{Chollet}. To speed up the convergence, we apply a batch--normalization\cite{szegedy2016inception, simard1998transformation}, setting the mean to zero and the variance to one in the input layers~\cite{ioffe2015batch, lecun1998gradient}. We found that this normalization not only makes the training faster but also improves the classification results, reaching an accuracy of 0.97 in the test prediction.


\subsection{M--RNN and AM--RNN optimization} \label{Sec:Optimization_Iupac}

Because of the analogies between \gls{M-RNN-AT} and \gls{AMRNN}, their loss functions and trainings are similar, hence we define the optimization problem for both at the same time with the notation used in the main text. The function to be maximized is the probability of obtaining a correct sentence $S = (S_1,...,S_N)$ given an input $(I, S_0')$:
\begin{equation}
\theta^* = arg\max_\theta \sum_{(I,S)}\log p(S|I, S_0';\theta)
\end{equation}
where $\theta$ represents the model parameters, and $I$ is the 3D image stack. Note that the prediction of the \gls{IUPAC} nomenclature, according to the decomposition performed on a set of terms, must depend on the predictions already performed, i.e. the prediction must take into account previously predicted terms in order to have semantic meaning, so it is a time series. 
During the training, we feed the model at each time step $t$ with the target of previous time steps $(S_0',S_1,...,S_{t-1})$ and not with the predictions performed $(Y_1,...,Y_{t-1}),$ which in some cases are wrong. We refer to each input of the model at the time step $t$ as the pair $(S_0',S_1...,S_{t-1};I).$ Note that, in this way, the final prediction length of each molecule depends on $t$. Thus, the prediction of $S$ depends on the prediction of each specific term, which in turn depends not only on $I$ but also on all the predictions performed in previous time steps. Since the model predicts a single term of the sequence at each time step, it is natural to apply the chain rule to model the joint probability over the sequential \textit{terms}. Hence, the probability of obtaining a correct prediction for the complete sequence is 
described by the sum of the logarithmic probabilities over the terms. 
Therefore, the maximum log--likelihood function is as follows:
\begin{equation}
L(S,I)=\sum_{t=1}^{N}\log p(S_t|I,S_0', S_1,...,S_{t-1};\theta).
\end{equation}
As the deep learning optimization techniques consist of searching for a minimum rather than a maximum, the loss function is described by the sum of the negative log--likelihood:
\begin{equation}
L(S,I)=-\sum_{t=1}^{N}\log p(S_t|I,S_0', S_1,...,S_{t-1};\theta).
\end{equation}

\subsection{M--RNN and AM--RNN training} \label{Sec:Training_Stages}

\begin{figure}[t!]
\centering
\includegraphics[clip=true, width=1.0\columnwidth]{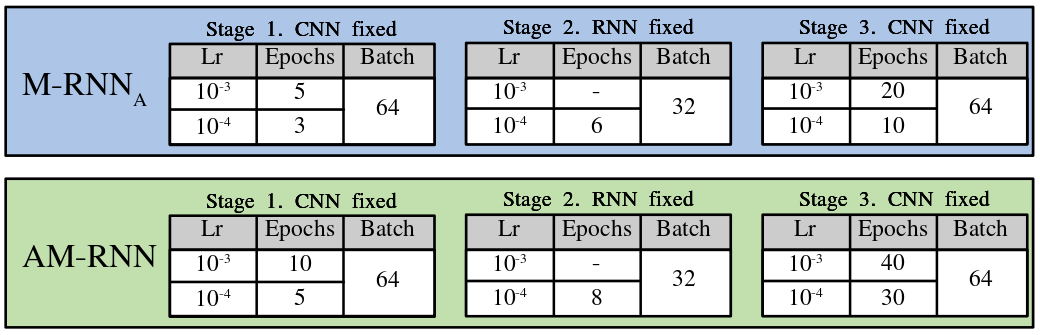}
\caption{Details of the training scheme for each stage of each of the models. Lr is a short for learning rate.}
\label{Fig:Training_details}
\end{figure}

The training of both \gls{M-RNN-AT} and \gls{AMRNN} proceeds in three stages in which the weights of the \gls{CNN} and RNN components are fixed (non-trainable) alternatively. In the first stage, the model initialises all its weights randomly except those of the \gls{CNN} component, which are pre--trained with the chemical classification explained in \cref{Sec:Class_Inception_ResNet}, and focuses on the training of the RNN. Although specialised in a different classification, the \gls{CNN} component output already represents high--level features of each input stack.
%
The model is then fed with the input $(I,S).$ 
In the same way as we do for the \gls{CNN} pre-training described in \cref{Sec:Class_Inception_ResNet}, a random combination of the simulation parameters described in \cite{QUAM-AFM_repository} is selected for each input at each epoch. Thus, although the \gls{CNN}'s weights are fixed, the high--level representation of each structure is different in each epoch.  This selection, coupled with the dropout layer of the $\varphi$ component, ensures that the RNN component does not overfit for specific representations of the input images. 

The aim of the second stage is to specialise the weights of the \gls{CNN} component in the semantic prediction. To this end, the weights of the RNN component are fixed and the \gls{IDG} described in \cref{Sec:Class_Inception_ResNet} is applied to the input stack. Furthermore, the selection of simulation parameters is randomly chosen for each input $I$. In this stage the prediction is performed for a single time step of each pair $(S,I).$
%
After completion of this second stage, the \gls{CNN} component provides specific details for the \gls{IUPAC} formulation rather than to the chemical classification. Finally,  the third training stage repeats the process of the first one, fixing the weights of the \gls{CNN}. 
Further details for the training at each stage (number of epochs, batch size, learning rate) are shown in \cref{Fig:Training_details}.

\newpage

\section{Influence of the molecular torsion in the model performance: gas--phase versus flat configurations} \label{Sec:Torsion_vs_planar}


HR-AFM shows an outstanding lateral resolution for quasi-planar molecules, but it is more limited to discern correctly molecules with a significant torsion. This is due to the nature of the contrast and the probe flexibility. The exponential behavior of the Pauli repulsion with respect to the tip-sample distance makes this contribution the most relevant contrast source. Thus, each atom in the sample will behave as an umbrella of around 2-3~{\AA} veiling any other electronic charge density below this region (except some cases where deformations in the charge density may occur). Furthermore, the CO molecule will tilt under this repulsion and will block the access of the probe to the region underneath. 

\begin{figure}[!b]
\centering
\includegraphics[clip=true, width=0.99\columnwidth]{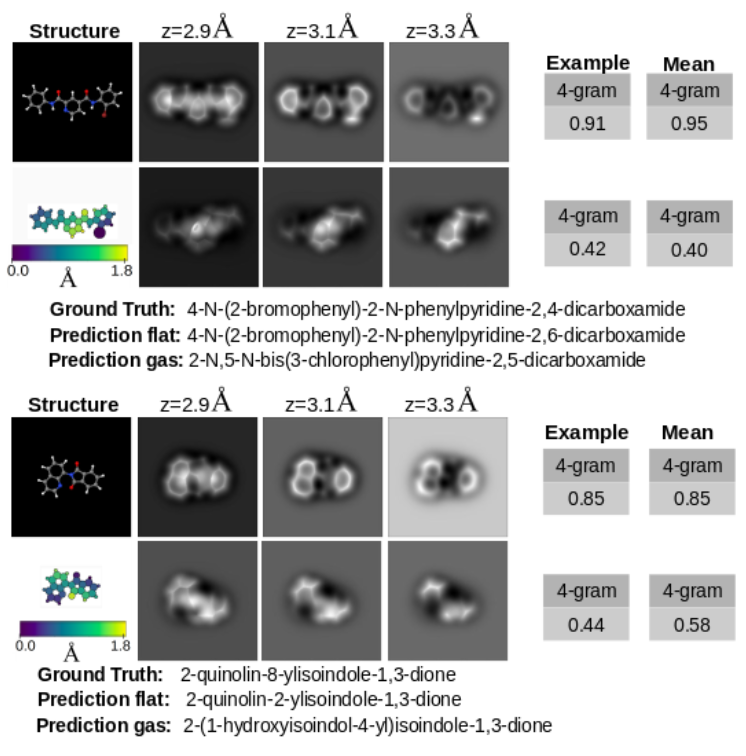}
 \caption{Comparison of the model predictions for the gas--phase and the forced planar configuration of a representative molecule, whose  ball--and--stick depiction and height map are shown on the left. The upper \gls{AFM} images correspond to the simulation with the structure in a completely planar configuration while the lower ones correspond to the images for the gas--phase structure. To the right of the \gls{AFM} images is the 4--gram score corresponding to the prediction shown (flat and gas--phase structure, respectively). Further to the right is the average of the 4--gram scores obtained for images generated with the 24 combinations of operational parameters considered in \gls{AFMD}. The prediction example shown is the one that scored closest to the mean.}
\label{Fig:Torsion_vs_planar1}
\end{figure}
\begin{figure}[b!]
\centering
\includegraphics[clip=true, width=0.99\columnwidth]{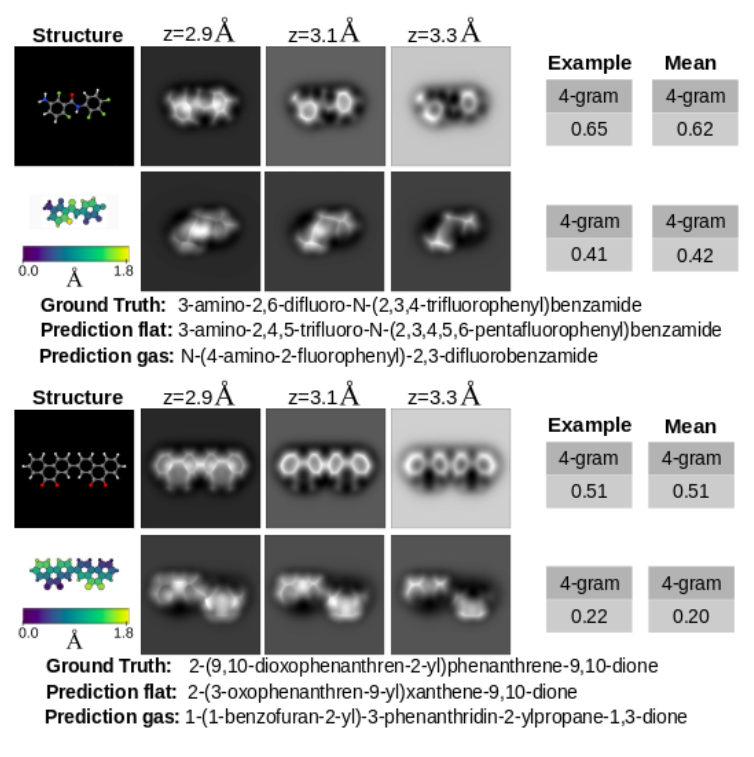}
\caption{As \cref{Fig:Torsion_vs_planar1} for two molecules that include chemical chemical groups, like F atoms (top) and diketones  (bottom), that display faint AFM features and thus are more difficult to recognise by the model.}
\label{Fig:Torsion_vs_planar2}
\end{figure}

Previous works\cite{alldritt2020automated} suggest that it is difficult to retrieve information from parts of the molecule that are located more than 1.5~\AA\ below the topmost atoms. According to this, we expect our algorithm to provide a poorer performance when finding out the IUPAC names of molecules showing significant out-of-plane distortions.
In order to quantify this, we have carried out some tests to directly show how the model accuracy improves for planar structures. We have selected four molecules showing a torsion of about 1.8~{\AA} and recalculated them in a forced planar configuration. 
These planar configurations have been determined by DFT calculations where we fix the $z$-position of all the atoms in the molecule and let them move in the $xy$ plane in order to reach the ground state configuration compatible with this constraint.
If we feed the algorithm with the image stacks of the planar configuration the \gls{BLEU} 4-gram score increases significantly in all cases,  even when the chemical nature of the molecule makes the recognition by our model more difficult. \Cref{Fig:Torsion_vs_planar1} shows two molecules where the prediction for the non planar failed but where we obtain  a highly accurate result (an almost perfect match in the cases shown) for the planar configurations, with  only an error in the misplacement of one of the functional groups.  \Cref{Fig:Torsion_vs_planar2} shows the result for two other molecules that contain functional groups that are difficult to discern such as fluorine atoms, easily mistakable with other terminations like diketones which may display faint AFM features. Although the \gls{BLEU} 4-gram for the planar configuration is not high there is a clear improvement compared to the non--planar cases.         

\clearpage

\section{Experimental test}

\begin{figure}[b!]
\centering
\includegraphics[clip=true, width=0.8\columnwidth]{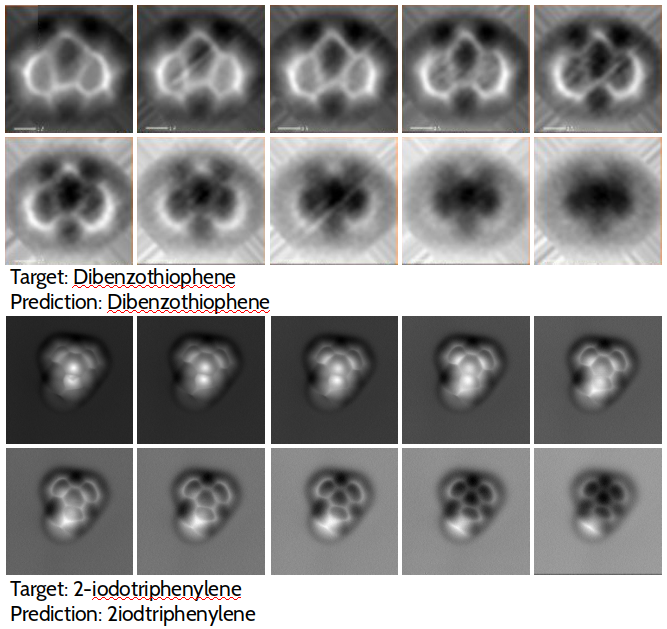}
 \caption{Experimental images used to test the molecular identification with our model. The top set of images corresponds to the 10 images of dibenzothiophene taken at 10 different tip-sample distances (covering a height range of 100pm) in ref.~\cite{EXP} (reproduced courtesy of the American Chemical Society, ACS), using a bias voltage $V=40$~mV and a tip oscillation amplitude of
approximately 50 pm. Images were aligned and Laplace filtered in the original publication to highlight the molecular structure while maintaining the feature shapes as well as possible.
The bottom set corresponds to 10 constant--height frequency shift images of 2-iodotriphenylene, covering a height range of 72 pm (with a height variation of 8 pm between two consecutive images), published in the supplementary material of ref.~\cite{martin2019bond} (reproduced courtesy of the American Physical Society, APS). A bias voltage $V= -0.57$~mV and an oscillation amplitude of $A = 52$~pm have been used for the measurements.}
\label{Fig:Experimental_Test}
\end{figure}
We have performed a test with 3D stacks of experimental AFM images for dibenzothiophene  and  2-iodotriphenylene (see~\cref{Fig:Experimental_Test}). In the first case, despite the strong noise in the images and the white lines crossing the images diagonally, we obtain a perfect prediction. In the second case, with images covering a tip-sample distance range of 72~pm (smaller that the 100 pm range used for the training), the prediction performed by the model is ``2iodtriphenylene", which is very close to the ground truth, just missing a hyphen but  providing all the relevant chemical information.  
Notice that, in both cases, experimental images have been scaled close to the size of the unit cell used in \gls{AFMD} before feeding the model. Without this scaling, the score in the 4-gram evaluation falls dramatically for both cases, suggesting that the typical sizes of chemical moieties learned during the training are an important component in the success of the identification process. This  sensitivity to variations in image sizes can possibly be reduced by increasing the zoom range of $\pm 15\%$ applied in our data augmentation during the training.

\newpage
\section{Embedding} \label{sec:embedding}

\begin{figure}[b!]
\centering
\includegraphics[clip=true, width=0.85\columnwidth]{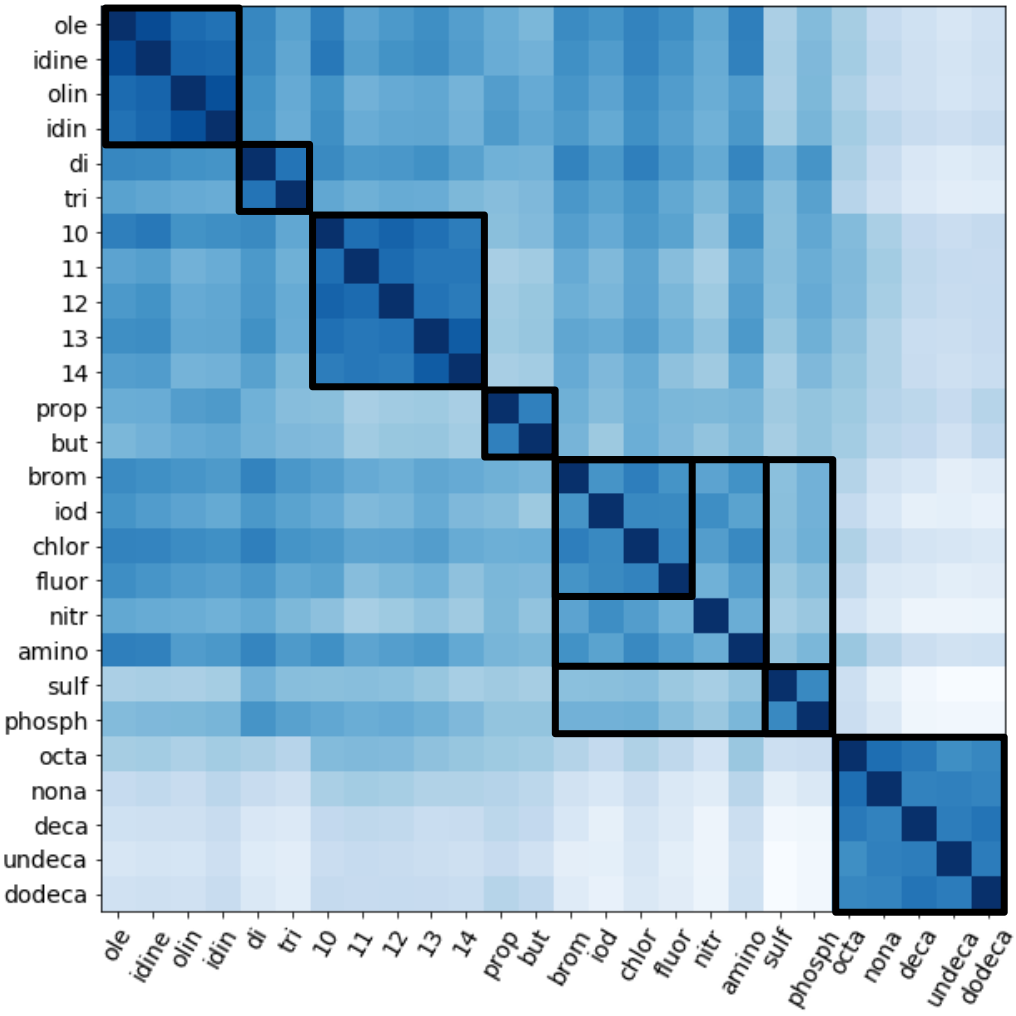}
\caption{Distances in the L2 sense between some of the terms in the \gls{AMRNN} embedding layer. The higher intensity of blue corresponds to smaller distances while the lighter ones correspond to longer distances. The distance from a term to itself is zero, so it matches the darkest blue.}
\label{Fig:Embedding}
\end{figure}

The weights of the embeddings of the RNN components of models applied to image captioning \cite{you2016image, vinyals2015show} are usually pretrained with Word Embeding to represent the semantic high-level features in a vectorial space. 
The usefulness of this technique lies in the representation of words in a vector space, in which words with similar semantic meaning are represented close together, as shown in \cref{Fig:Embedding}. In this way, the models manage to use synonyms providing versatility and improving its expressive capacity. The versatility of languages to express the same idea in different ways is further exploited in image captioning providing the same input image with different annotation outputs. Then those models succeed in learning that different words have similar meaning. 

However, there are no trivial synonyms in our case, due to the close relationship between the IUPAC names assigned to the functional groups in a molecule. Thus, any change in the output sequence will result likely in an error, as a different molecule would be predicted. 
Additionally, there is no freely available pre-trained Word Embeding for the \gls{IUPAC} nomenclature, and therefore all weights of the RNN component are initialised randomly. Consequently, we train the embedding layer with the rest of the RNN component (stages 1 and 3 described in \cref{Sec:Training_Stages}). Here we find that, although the \gls{IUPAC} nomenclature does not use synonyms, the representations made in this vector space have certain semantic relationships. To check this, we define the distance in the L2 sense and calculate the distances among the terms (of the \gls{AMRNN}). We find, for example, that the terms represented closest to {\em brom} are {\em chlor, fluor} and {\em iod}. In addition to the terms associated with halogens, those designating numbers are also represented in close proximity. It is also noteworthy that the terms closest to {\em nona} are  {\em octa, deca, undeca} and {\em dodeca}, as shown in \cref{Fig:Embedding}. It should be recalled that each of the terms is represented by a number to feed the network, so the model is not learning lexical relations but syntactic relations.




\newpage
\bibliographystyle{Science}
\bibliography{supplemental}
\addcontentsline{toc}{section}{References and Notes}